\documentclass[eqsecnum,aps,amssymb]{revtex4}

\def\e{\varepsilon}
\def\th{\vartheta}

\def\r{\rho}

\def\oo{\omega}
\def\Th{\Theta}
\def\O{\Omega}

\def\square{{\vcenter{\vbox{\hrule height.4pt \hbox{\vrule width.4pt
height1.45ex \kern1.45ex \vrule width.4pt}
\hrule height.4pt}}}}

\def\qed{\penalty10000\hfill$\square$\par\goodbreak\medskip}

\def\E{{\cal E}}
\def\H{{\cal H}}

\def\PP{{\mathbb P}}
\def\EE{{\mathbb E}}

\def\tr{{\rm tr}\,}

\def\half{\textstyle{1\over2}}

\def\tuple#1_#2{#1_1,#1_2,\ldots,#1_{#2}}
\def\tup#1_#2{#1_1,\ldots,#1_{#2}}

\def\dd#1{{d \over d#1}}

\def\Nti#1{\widetilde{N}^{i}_{#1}}
\def\Wti#1{\widetilde{W}^{i}_{#1}}
\def\Wtj#1{\widetilde{W}^{j}_{#1}}

\newenvironment{procl}[1]{\bigskip\begin{sloppypar}
                          \noindent{\bf #1.}\sl}
                         {\hfill\end{sloppypar}}

\newenvironment{proof}{\bigskip\begin{sloppypar}\noindent{\bf Proof.}}
                      {\hfill\qed\end{sloppypar}}

\begin{document}
\draft

\title{A Pathwise Ergodic Theorem for Quantum Trajectories}

\author{B.~K\"ummerer$^1$ and H.~Maassen$^2$}
\address{$^1$Fachbereich Mathematik,
         Technische Universit\"at Darmstadt,
         Schlossgartenstra\ss e 7,
         D-64 289 Darmstadt
         Germany,
         {\tt kuemmerer@mathematik.tu-darmstadt.de}\\
         $^2$Mathematisch Instituut,
         Radboud Universiteit Nijmegen,
         Toernooiveld 1,
         6525 ED Nijmegen,
         The Netherlands,
         {\tt maassen@math.kun.nl}}

\date{June 25, 2004}
\begin{abstract}
\noindent
If the time evolution of an open quantum system approaches
equilibrium in the time mean, then on any single trajectory of
any of its unravelings the time averaged state approaches the
same equilibrium state with probability 1.
In the case of multiple equilibrium states the quantum trajectory
converges in the mean to a random choice from these states.
\end{abstract}

\maketitle

\section{Introduction}

\noindent
Stochastic Schr\"odinger equations and their solutions, quantum
trajectories, have been extensively studied in the last 15 years
(cf. \cite{Carmichael}, \cite{GardinerZoller}).
They provide insight into the behaviour of open quantum systems
and they are invaluable for Monte Carlo
simulations of the time evolution of such systems,
in particular for the numerical determination of equilibrium states.

In performing such simulations one is confronted with the problem,
whether it is necessary to average over many trajectories,
or if it suffices to calculate the time average over a single trajectory,
which is often more convenient (cf. \cite{GardinerZoller}).

In this paper we prove that for any finite-dimensional quantum system
and for any initial state the time average of a single quantum
trajectory converges to some equilibrium state with probability one.
This result holds true despite the fact that the quantum
trajectory itself may stay away from equilibrium forever.

In the simple case that there exists only one equilibrium state,
the above result implies that the path average converges to this
particular state,
almost surely and independently of the starting point chosen.
So in one sense the quantum trajectory is ergodic in this case:
The path average of any observable of the quantum system equals
its expectation in the equilibrium state.
However,
when looked upon as a classical stochastic process with values
in the space of all quantum states,
the quantum trajectory need not be ergodic,
even in this simple and well-behaved case:
There may be disjoint regions in the space of all quantum states
between which no transitions are possible.

In a previous paper \cite{KueMaaErg1} we have considered the
ergodic properties of the observed output of open quantum systems.
We found that quantum systems with finite dimensional Hilbert spaces
and unique equilibrium states lead to ergodic observations.
Strangely enough,
the techniques needed to prove our present result seem to be entirely
different from the ones used in that paper.
Here we make strong use of martingales,
which have been introduced to this context in \cite{Belavkin}.
As in \cite{KueMaaErg1} we concentrate our discussion
on jump processes in continuous time using the formulation of Davies
and Srivinas \cite{Davies} \cite{SrinivasD}.
But the result also holds for diffusive Schr\"odinger equations and for
quantum evolutions in discrete time, as they occur in repeated measurement
situations like the micromaser \cite{WeBuKuMa}.

The paper is organizied as follows. We formulate our result in
Section II and introduce the necessary martingales in Section III.
In Section IV the proof of the theorem is given.
It is extended to the diffusive and discrete time cases
in Sections V and VI, respectively.


\section{The main result}

\noindent
The state of an open quantum system is described by a density matrix $\rho$ on
a finite dimensional Hilbert space $\H$, obeying a Master equation
$\dot\rho=L\rho$, where $L$ is a generator of Lindblad form
\cite{Lindblad},
   $$L(\r)=i[H,\r]+\sum_{i=1}^k V_i\r V_i^*-\half(V_i^*V_i\r- \r
   V_i^*V_i)\;.$$
Here $H,V_1,\ldots,V_k$ are linear operators on $\H$,
$H$ being self-adjoint.

Conservation of normalisation of $\rho$ is expressed by the relation

\begin{equation}
\tr L(\rho)=0\quad\hbox{for all }\rho\;.
\label{normal}
\end{equation}
An unraveling of $\rho$ is induced by a decomposition of the generator
\begin{equation}
   L=L_0+\sum_{i=1}^k J_i\;,
\label{unravel}
\end{equation}
where $J_i(\r )=V_i\r V_i^*$ would be a natural choice.
In general, any decomposition can be treated for which $e^{tL_0}$, $t\ge0$,
and $J_i$, $i=1,\ldots,k$, are completely positive.

This decomposition may be interpreted as follows.
The open system is under continuous observation by use of $k$ detectors.
The reaction of the detectors to the system consists
of clicks at random times.
The evolution $\rho\mapsto e^{tL_0}(\rho)$ denotes the change of the
state of the system under the condition that during a time interval
of length $t$ no clicks are recorded.
The operator $\rho\mapsto J_i(\rho)$ on the state space
describes the change of state conditioned on the occurrence of a click
of detector $i$.

So, if $\rho$ describes the state of the system at time 0, and if,
during the time interval $[0,t]$, clicks are recorded at times $\tuple
t_n$ of detectors $\tuple i_n$ respectively, and none more, then, up
to normalisation, the state at time $t$ is given by

\begin{equation}
\th_t((t_1,i_1), \ldots , (t_n,i_n) ) =
e^{(t-t_n)L_0} J_{i_n} e^{(t_n-t_{n-1})L_0}\cdots e^{(t_2-t_1)L_0}
                       J_{i_1} e^{t_1L_0}(\rho)\;.
\label{probab}
\end{equation}
The probability density
for these clicks to occur is equal to the trace of
$\th_t$ in (\ref{probab}). We shall denote the normalized density
matrix $\th_t/\tr(\th_t)$ by $\Theta_t$.

We imagine the experiment to continue indefinitely. The observation
process will then produce an infinite detection record
$\bigl((t_1,i_1),\ (t_2,i_2),\ (t_3,i_3),\ \dots\bigr)$\ ,
where we assume that
$0 \le t_1 \le t_2 \le t_3 \le \dots$\ , and
$\lim_{n\to\infty} t_n = \infty$ (i.e., the clicks do not accumulate).
Let $\Omega$ denote the space of all such detection records
with Lebesgue-measure
$$d\oo =
 \sum_{n=0}^\infty \ \ \  \sum_{i_1 = 1}^k \cdots \sum_{i_n = 1}^k dt_1\cdots
 dt_n  \ . $$

As was shown in \cite{KueMaaErg1}, each initial state
$\th_0$ determines a probability measure $\PP^{\th_0}$ on
$\O$ whose restriction to the time interval $[0,t]$ has density
$\tr(\th_t)$ as described above.
We may
consider $(\Theta_t)_{t \ge 0}$ as a stochastic process on this
probability space taking values in the density matrices.
A path of this process is called a {\sl quantum trajectory}.
We thus obtain an unraveling of the state at time $t\ge0$:

\begin{equation}
   T_t(\th_0):= e^{tL}(\th_0)=
       \int_\O \Theta_t(\oo)\PP^{\th_0}(d\oo)=\EE^{\th_0}(\Theta_t)\;.
\label{pathunravel}
\end{equation}

So far the framework is essentially the same as described in
our previous paper \cite{KueMaaErg1}.
It is the framework frequently used in computer simulations
(cf., e.g., \cite{Carmichael}, \cite{GardinerZoller}).
If one is only interested in
the average evolution $e^{tL}$, then the decompositon \ref{unravel}
can be chosen at will.

We now address the question, what can be said about the asymptotic
behaviour of each single quantum trajectory $(\Th_t(\oo))_{t\ge0}$.

Let us denote by $\E$ the space of equlibrium states,
i.e. density matrices $\rho$ which
are left invariant by the average evolution $T_t$.
Since the Hilbert space is finite dimensional the limit

\begin{equation}
   P(\th) = \lim_{t \to \infty} \frac{1}{t} \int_0^t T_s(\th) ds
\label{meanprojection}
\end{equation}
exists and projects any density matrix $\th$ onto the space $\E$ of
equilibrium states.


\begin{procl}{Theorem 1}
Suppose that $T_t = e^{tL}$ has only a single equilibrium state
$\rho$. Then for every initial state $\th_0$ the
quantum trajectory $(\Theta_t)_{t \ge 0}$ satisfies

$$ \lim_{t \to \infty} \frac{1}{t} \int_0^t \Theta_s(\oo)ds = \rho\ ,$$
for almost all $\oo$ with respect to the probability measure $\PP^{\th_0}$.

More generally, in the case that there is more than one equilibrium
state, one has almost surely

$$ \lim_{t \to \infty} \frac{1}{t} \int_0^t \Theta_s(\oo)ds
                                            = \Theta_\infty(\oo)\ ,$$
where $\Theta_\infty$ is a random variable, depending on the initial
state $\th_0$, and taking values in the equilibrium states.
The expectation of $\Th_\infty$ is $P(\th_0)$.
\end{procl}

\smallskip
The proof of this theorem is inspired by the arguments leading to
Breiman's strong law of large numbers for Markov chains \cite{Breiman}
(see also \cite{Krengel}), which however does not apply directly
to the situation of continuous time quantum trajectories.
Our proof, based on the martingale convergence theorem (Section III),
will be given in Section IV.
In our discussion we make free use of standard stochastic notation and arguments
for which we refer, e.g., to \cite{martingale}, \cite{van Kampen},
\cite{Chung Williams}.

\section{Martingales}

\noindent The process $(\Theta_t)_{t\ge 0}$ consists of smooth
evolution according to $e^{tL_0}$ interrupted by jumps of different
types $i=1,\ldots,k$, namely $\Th_t \mapsto
J_i\Th_t/\tr(J_i\Th_t)$. Let $N_i(t)$ denote the number of jumps of
type $i$ before time $t$.
In the theory of point processes
\cite{Ramakrishnan}, \cite{van Kampen}, \cite{Bartlett}
it is well known that,
from the probability density (\ref{probab}),
it follows that the unconditioned probability density of the occurrence
of a jump of type $i$ at time $u$,
given the state $\Th_s$ at time $s$ is
   $$\tr\bigl(T_{t-u}J_iT_{u-s}(\Th_s)\bigr)
     =\tr\bigl(J_iT_{u-s}(\Th_s)\bigr)\;$$
for $t \ge u \ge s \ge 0$, independent of $t \ge u$.
Let $\EE^{\th_0}_s$ denote expectation with respect to $\PP^{\th_0}$,
given the process up to time $s$.

In a similar way as (\ref{pathunravel}) follows from (\ref{probab}),
it is easy to show that $T_{u-s}(\Theta_s) = \EE_s^{\th_0}(\Theta_u)$,
and therefore

$$\EE^{\th_0}_s\left(N_t^i - N^i_s\right)
   = \int_s^t \tr\bigl(T_{t-u}J_iT_{u-s}(\Th_s)\bigr)\; du
   = \EE^{\th_0}_s\left(\int_s^t \tr\bigl(J_i(\Th_u)\bigr)du \right)\ .$$
If we now denote by $\tilde N_t^i$ the process

$$\tilde N_t^i := N_t^i - \int_0^t \tr J_i(\Th_u)du\ ,$$
then $\tilde N_t^i$ is a {\sl martingale}, 
\cite{martingale}, i.e., for all
$0 \le s \le t$:

  $$\EE^{\th_0}_s(\tilde N_t^i) = \tilde N_s^i\ .$$
$(\tilde N_t^i)_{t \ge 0}$ is the {\sl compensated
number process} of jumps of type $i$.

\begin{procl}{Lemma 2}
The quantum trajectory $(\Th_t)_{t \ge 0}$ satisfies the stochastic
Schr\"odinger equation \cite{Carmichael}, \cite{BoGuMa}

\begin{equation}
d\Th_t = L(\Th_t)dt
          + \sum_{i=1}^k \left(\frac{J_i(\Th_t)}{\tr(J_i(\Th_t))}-\Th_t \right)
               d\tilde N^i_t \ .
\label{stochschr}
\end{equation}
where the stochastic differential equation is interpreted
in the sense of It\^o \cite{Chung Williams}.

\end{procl}

\begin{proof}
Between jumps
$\th_t$ evolves according to $\dd t \th_t=L_0(\th_t)$,
at a jump of type $i$ it jumps from $\th_t$ to $J_i(\th_t)$.
It follows that the normalised state $\Th_t=\th_t/\tr(\th_t)$
satisfies
$$d\Th_t=\dd t\left({{\th_t}\over{\tr(\th_t)}}\right)\,dt
   +\sum_{i=1}^k\left({{J_i(\Th_t)}\over{\tr(J_i(\Th_t))}}-\Th_t\right)
       \,dN^i_t\;.$$
Since between jumps we have
$$
\dd t\left( {\th_t \over \tr(\th_t)}\right)
 = {L_0(\th_t) \over \tr(\th_t)} -
   {\th_t \cdot \tr(L_0(\th_t)) \over \tr(\th_t)^2}
 = L_0(\Th_t) - \Th_t\cdot \tr(L_0(\Th_t))
$$
and since $dN^i_t = d\Nti t + \tr(J_i(\Th_t))dt$,
we have, using that $\tr \circ L = 0$,

\begin{eqnarray*}
d\Th_t &= &\Bigl(L_0(\Th_t)- \Th_t\cdot \tr L_0(\Th_t)\Bigr) dt
         + \sum_{i=1}^k\left({{J_i(\Th_t)}\over{\tr(J_i(\Th_t))}}-\Th_t\right)
           \cdot \Bigl(d\Nti t + \tr(J_i(\Th_t)) dt\Bigr)\\
       &= &\left(L_0 + \sum_{i=1}^k J_i\right)(\Th_t)\; dt -
         \Th_t\cdot \tr\left( \biggl(L_0 + \sum_{i=1}^k
         J_i\biggr)(\Th_t)\right)dt
        + \sum_{i=1}^k\left({{J_i(\Th_t)}\over{\tr(J_i(\Th_t))}}-\Th_t\right)
           \cdot d\Nti t\\
       &= &L(\Th_t)dt
          + \sum_{i=1}^k \left(\frac{J_i(\Th_t)}{\tr(J_i(\Th_t))}-\Th_t \right)
               d\tilde N^i_t \ .
\end{eqnarray*}

\end{proof}

\noindent
The process $\Theta_t$ starts at $\Theta_0=\th_0$.
Let us now consider two other stochastic processes
\begin{eqnarray*}
M_t &:= &\Th_t - \th_0 - \int_0^t L(\Th_s) ds
      =  \int_0^t \sum_{i=1}^k \left(\frac{J_i(\Th_s)}
                                           {\tr J_i(\Th_s)}
                       -      \Th_s  \right)d\tilde N^i_s\ \ (t \ge 0)\ , \\
      &\mbox{and}&\\
Y_t &:= &\int_1^t \frac1s \sum_{i=1}^k \left(\frac{J_i(\Th_s)}
                                           {\tr J_i(\Th_s)}
                       -      \Th_s  \right)d\tilde N^i_s
                        =\int_1^t \frac 1s\; dM_s \ \ (t \ge 1) \ .
\end{eqnarray*}

\noindent
From the fact that $\Nti t$ is a martingale, it follows that
these processes are martingales as well \cite{Chung Williams}.


\noindent
We now come to the main result of this section

\begin{procl}{Proposition 3}
For any intial state $\th_0$ the quantum trajectory
$(\Th_t(\oo))_{t \ge 0}$ satisfies
$$ \lim_{t \to \infty} \frac 1t \int_0^t L(\Th_s(\oo))ds = 0$$
almost surely with respect to $\PP^{\th_0}$.
\end{procl}

\begin{proof} Let us first consider the martingale $Y_t$ which takes
values in the self-adjoint matrices. In order to conclude from the
martingale convergence theorem \cite{martingale} that $(Y_t)_{t\ge 0}$
converges almost surely, we show that $\EE^{\th_0}(\tr(Y_t^2))$
remains bounded:

Denote the coefficient
$\left( \frac{J_i(\Th_t)}{\tr J_i(\Th_t)} - \Th_t \right)$
by $X_t^i$. Then
$$ dY_t = \sum_{i=1}^k {1 \over t}X_t^id\Nti t\ .$$
By the It\^o rules for jump processes \cite{Chung Williams}
$d\tilde N^i_t d\tilde N^j_t = d N^i_t d N^j_t =  \delta_{ij} dN^i_t$,
we find that

$$(dY_t)^2 = \frac1{t^2}\sum_{i=1}^k \sum_{j=1}^k X^i_t X^j_t
                         d\tilde N^i_t d\tilde N_t^j
           = \frac1{t^2}\sum_{i=1}^k (X^i_t)^2 dN^i_t \ .$$

From
$d(Y_t^2) = 2Y_tdY_t + (d Y_t)^2$
and $\EE^{\th_0}(d\tilde N_t^i) = 0$, hence
$\EE^{\th_0}(\tr (Y_t dY_t)) = 0$,
we obtain
$\EE^{\th_0}(d(\tr(Y_t^2))) = \EE^{\th_0}\bigl(\tr((d Y_t)^2)\bigr)$.
Therefore, since $\EE^{\theta_0}(dN^i_t) = \EE^{\theta_0}(\tr J_i(\Th_t))dt$,
$$
d\EE^{\th_0}\bigl(\tr(Y_t^2)\bigr) = \EE^{\th_0}\bigl(\tr((d Y_t)^2)\bigr)
= \frac1{t^2}\sum_{i=1}^k \EE^{\th_0}\left(\tr((X^i_t)^2)\cdot
\tr(J_i(\Th_t))\right) dt
$$
hence
\begin{eqnarray*}
\EE^{\th_0}\bigl(\tr (Y_t^2)\bigr)
   &= &\int_1^t \frac1{s^2}
          \sum_{i=1}^k \EE^{\th_0}\left( \tr((X_s^i)^2)
             \cdot \tr J_i(\Th_s)\right)ds
   \le  4 \sum_{i=1}^k \Vert J_i\Vert \ .
\end{eqnarray*}

\noindent
In this sense, $(Y_t)_{t \ge 1}$
is $L^2$-bounded and it
follows that $Y_t$ converges almost surely to some
random variable $Y$.
In particular, since $Y_t$ is continuous up to finitely many
jumps on compact time intervals and has a limit as $t \to \infty$
almost surely, it is bounded almost surely.
Therefore, applying the partial integration formula, which is
also valid if $Y_t$ has jumps, we obtain for $t \ge 1$

$$
M_t =  M_1 + \int_1^t s\;dY_s
    =  M_1 + sY_s\Big|_1^t - \int_1^t Y_s\; ds
    =  M_1 + tY_t - \int_1^t Y_s\; ds \ ,
$$
therefore,
\begin{eqnarray*}
\lim_{t \to \infty} \frac 1t M_t
                &= & \lim_{t \to \infty} \frac 1t M_1 +
                \lim_{t \to \infty} Y_t -
                \lim_{t \to \infty}\frac 1t \int_1^t Y_s\; ds\\
                &= & 0 + Y - Y\\
                &= & 0 \ .
\end{eqnarray*}
We thus conclude that
$$\lim_{t \to \infty} \frac 1t
    \left( \Th_t - \th_0 - \int_0^t L(\Th_s)\;ds \right) = 0\ .$$
As ($\Th_t - \th_0$) remains bounded, the statement of the Proposition
follows.
\end{proof}


\section{Proof of the main result}

\noindent
We shall prove Theorem 1 in two steps.

\begin{procl}{Step 1}
If $P$ is given as in \ref{meanprojection}, then
for any initial state $\th_0$ the limit
$$\lim_{t \to \infty} P(\Th_t) =: \Th_\infty$$
exists almost surely with resprect to $\PP^{\th_0}$, and
satisfies $\EE^{\th_0}(\Th_\infty) = P(\th_0)$.
\end{procl}

\begin{proof}
Acting with the operator $P$ on both sides of
(\ref{stochschr}) in Lemma 2 we see
that $\EE^{\th_0}\bigl(P(d\Th_t)\bigr) = 0$, hence
$(P(\Th_t))_{t \ge 0}$
is a martingale. Since it takes values
in the states it is bounded, and therefore
it converges almost surely, say to the random variable
$\Th_\infty$. The expectation of $\Th_\infty$ is
$P(\th_0)$, the initial value of the martingale $(P(\Th_t))_{t\ge 0}$.
\end{proof}

\begin{procl}{Step 2}
For any initial state $\th_0$ we have, almost surely with respect
to $\PP^{\th_0}$:

\begin{equation}
\lim_{t \to \infty} \frac 1t \int_0^t(\Th_u - P(\Th_u))\; du = 0\ .
\label{step3}
\end{equation}
\end{procl}

\begin{proof}
First we show that, for all $s\ge 0$,

\begin{equation}
\lim_{t \to \infty} \frac 1t \int_0^t(\Th_u - T_s(\Th_u))\; du = 0\ .
\label{step3pr}
\end{equation}

\noindent
Indeed, since $\frac d{dv}T_v = T_vL$ :
\begin{eqnarray*}
\int_0^t(T_s - id)(\Th_u)du &= &\int_0^t\int_0^s T_vL(\Th_u)\; dvdu\\
           &= &\int_0^s T_v\left( \int_0^t L(\Th_u)\;du \right) dv\ .
\end{eqnarray*}
Dividing by $t$ and taking the limit $t \to \infty$, we obtain
\ref{step3pr} by Proposition 3.

Clearly, averaging \ref{step3pr} over $[0,s]$ preserves its
validity:

$$
\lim_{t \to \infty} \frac 1t \int_0^t
 \left(\Th_u - \frac 1s \int_0^s T_v(\Th_u)\; dv\right)\; du = 0\ .
$$

\noindent
In the above we want to take the limit $s \to \infty$
{\sl before the limit $t \to \infty$}, in order to obtain the
statement \ref{step3} to be proved:

This is allowed since $\H$ is finite-dimensional: Then for $\e > 0$
there exists $s > 0$ such that
$\big\Vert \frac 1s \int_0^s T_v dv - P  \big\Vert < \frac\e 2 \ , $
hence
$$\Big\Vert \frac 1s \int_0^s T_v(\Th_u(\oo)) dv - P(\Th_u(\oo))  \Big\Vert <
\frac \e 2 \ , $$
uniformly in $u$. For $\PP^{\th_0}$-almost every $\oo \in \Omega$ we find
$t_0$ such that for $t > t_0$
$$\Big\Vert
 \frac 1t \int_0^t
 \left(\Th_u(\oo) - \frac 1s \int_0^s T_v(\Th_u(\oo))\; dv\right)\; du
\Big\Vert < \frac \e 2 \ .
$$

\noindent
Then, we obtain for such $t$
\begin{eqnarray*}
\Big\Vert \frac 1t
       \int_0^t(\Th_u(\oo) &-& P(\Th_u(\oo)))\; du\Big\Vert\\
    &=&  \Big\Vert
       \frac 1t \int_0^t\Bigl(\Th_u(\oo) - P(\Th_u(\oo))\
       + \frac 1s \int_0^s T_v(\Th_u(\oo)) dv
       - \frac 1s \int_0^s T_v(\Th_u(\oo)) dv \Bigr) \; du
          \Big\Vert \\
    &\le&  \Big\Vert
    \frac 1t \int_0^t
  \left(\Th_u(\oo) - \frac 1s \int_0^s T_v(\Th_u(\oo))\; dv\right)\; du
       \Big\Vert +
    \Big\Vert \frac 1t \int_0^t
         \Bigl(  \frac 1s \int_0^s T_v(\Th_u(\oo)) dv - P(\Th_u(\oo)) \Bigr) \; du
    \Big\Vert \\
   & < &\frac\e 2 + \frac \e 2 = \e \ .
\end{eqnarray*}

\end{proof}


\section{Diffusive quantum trajectorties}

\noindent
The ergodic result obtained above is not confined to jump processes.
Solutions of the Master equation $\dot\r=L\r$ with
   $$L(\r)=i[H,\r]+\sum_{j=1}^k V_j\r V_j^*-\half(V_j^*V_j\r-\r V_j^*V_j)$$
can alternatively be unraveled into a diffusion $\Th_t$ on the state space,
satisfying the stochastic differential equation [Bel], [Car], [BGM]
   $$d\Th_t=L(\Th_t)dt+\sum_{i=1}^k X^i_t\,d\Wti t\;,$$
where
\begin{eqnarray*}
X^i_t&=&\Th_t V_i^*+V_i\Th_t-
                    \tr\bigl(\Th_t V_i^*+V_i\Th_t\bigr)\cdot\Th_t
                    \mbox{\qquad and}\\
              d\Wti t &=&dW^i_t-\tr\bigl(\Th_tV_i^*+V_i\Th_t\bigr)\,dt\;.\\
\end{eqnarray*}

\noindent
As usual, $W^i_t$, $i=1,\ldots,k,$ denote pairwise independent real-valued
Wiener processes. In this situation our main theorem takes the
following form.

\begin{procl}{Theorem 4}
We have almost surely
    $$\lim_{t\to\infty}{1\over t}\int_0^t\Th_s(\oo)\,ds=\Th_\infty(\oo),$$
where $\Th_\infty$ is a random variable, depending on the initial
state $\th_0$ and taking values in the equilibrium states.
Again, the expectation of $\Th_\infty$ is $P\th_0$.
\end{procl}

\begin{proof}
We follow the same line of argument as for jump processes.
Here we only discuss the modifications needed for the diffusive case.
We consider the stochastic processes $(M_t)_{t\ge0}$ and $(Y_t)_{t\ge1}$
given by
\begin{eqnarray*}
M_t &:= &\Th_t - \th_0 - \int_0^t L(\Th_s) ds
      =  \int_0^t \sum_{i=1}^k X^i_s
                       d\tilde W^i_s\ , \\
\mbox{and}&&\\
Y_t &:= &\int_1^t \frac1s \sum_{i=1}^k  X^i_s
                       d\tilde N^i_s
         =\int_1^t \frac 1s\; dM_s     \ .
\end{eqnarray*}

As was shown in [Bel], [BGM], these are martingales.
Again $\EE^{\th_0}(\tr(Y_t^2))$ remains bounded.
Indeed $d\Wti t d\Wtj t=dW^i_t\,dW^j_t=dt$ by the It\^o rules
and
$\EE^{\th_0}(d(\tr Y_t^2)) = \EE^{\th_0}(\tr (dY_t)^2)$
with

$$(dY_t)^2 = \frac1{t^2}\sum_{i=1}^k \sum_{j=1}^k X^i_t X^j_t
                         d\tilde W^i_t d\tilde W_t^j
           = \frac1{t^2}\sum_{i=1}^k (X^i_t)^2 dt \ ,$$
so that

$$
\EE^{\th_0}(\tr Y_t^2)
   =  \int_1^t \frac1{s^2}
          \sum_{i=1}^k \EE^{\th_0}\left( \tr(X_s^i)^2\right) ds
   \le  4 \sum_{i=1}^k \Vert V_i\Vert^2 \ .
$$

The partial integration argument, which is also valid for diffusions,
leads to Proposition 3. Step 1 and Step 2
in the proof of the main result remain unchanged.

\end{proof}


\section{Quantum Trajectories in Discrete Time}

\noindent
Our ergodic theorem also has a natural version in discrete time.
Let us briefly sketch the setting. A time evolution in discrete time
is given by the powers of a completely positive operator $T$ with
$\tr \circ T = \tr$. A Kraus decomposition
$$ T(\rho) = \sum_{i=1}^k V_i \rho V_i^*$$
of $T$ leads to an unraveling of this time evolution:
Let $\Omega$ be the set of all infinite sequences
$(\oo_1, \oo_2, \dots)$ with $\oo_j = 1, \dots ,k$. An initial state
$\th_0$ induces a probability measure $\PP^{\th_0}$ on $\O$
which is uniquely determined by the condition
$$\PP^{\th_0}\bigl(\{\oo \in \O:
     \oo_1 = i_1, \oo_2 = i_2, \ldots ,\oo_n = i_n\}\bigr)
     = \tr(V_{i_n} \cdots V_{i_1}\, \th_0\, V^*_{i_1} \cdots V^*_{i_n}) \ .$$
Then an unraveling of the time evolution $(T^n)_{n \ge 0}$
is given by the Markov chain
$(\Th_n)_{n \ge 0}$ on $(\O,\PP^{\th_0})$ with
$$ \Th_n(\oo) =
   \frac{V_{i_n} \cdots V_{i_1}\, \th_0\, V^*_{i_1} \cdots V^*_{i_n}}
        {\tr(V_{i_n} \cdots V_{i_1}\, \th_0\, V^*_{i_1} \cdots
        V^*_{i_n})} \ .
$$

\begin{procl}{Theorem 5}
As $ N \to \infty $, the averaged process
   $$\frac 1N \sum_{n=0}^{N-1} \Th_n(\oo)$$
converges $\PP^{\th_0}$-almost surely to a random equilibrium state $\Th_\infty$
with expectation $P(\th_0)$.
\end{procl}

The proof is a discrete version of the argument in the previous
sections, which corresponds to a variation on Breiman's indiviual ergodic
theorem for Markov chains \cite{Breiman}, \cite{Krengel}.



\begin{references}

\bibitem[Bel]{Belavkin} V.~P. Belavkin, Commun. Math. Phys. 146
(1992), 611 - 635.

\bibitem[BGM]{BoGuMa} L.~Bouten, M.~Guta, H.~Maassen,
Journ. Phys. A: Math Gen. 37 (2004), 3189 - 3209.


\bibitem[Bre]{Breiman} L. Breiman, Annals of Mathem. Stat. 31 (1960),
801 - 803.

\bibitem[Car]{Carmichael}
H.~Carmichael, {\sl An open systems approach to quantum optics},
Berlin, Heidelberg, New York: Springer 1993.

\bibitem[Dav]{Davies}
E.B.~Davies,
{\sl Quantum theory of open systems},
Academic Press, New York 1976.

\bibitem[Doo]{martingale}
J.L. Doob, {\sl Stochastic Processes},
John Whiley \& Sons, New York 1953.


\bibitem[GaZ]{GardinerZoller} C.W. Gardiner, P. Zoller,
{\sl Quantum Noise}, Springer, Berlin-Heidelberg 2000.

\bibitem[Kra]{Kraus}
K.~Kraus,
Ann. Phys. {\bf 64}, 331--335 (1971).

\bibitem[Kre]{Krengel} U. Krengel, {\sl Ergodic Theorems},
Walter de Gruyter, Berlin 1985.

\bibitem[K\"uM]{KueMaaErg1}
B.~K\"ummerer, H.~Maassen, J. Phys. A: Math Gen. 36 (2003)
2155 - 2161.

\bibitem[Lin]{Lindblad}
G.~Lindblad,
Commun. Math. Phys. {\bf 48}, 119--130 (1976).

\bibitem[MaK]{MaassenK}
H.~Maassen, B.~K\"ummerer,
in: Mini-proceedings of the Workshop on Stochastics and Quantum
Physics, Maphysto Centre, Aarhus 1999.

\bibitem[SrD]{SrinivasD}
M.D.~Srinivas, E.B.~Davies,
Optica Acta {\bf 28}, 981--996 (1981).

\bibitem[vKa]{van Kampen}
N.G. van Kampen, {\sl Stochastic Processes in Physics and Chemistry},
North Holland, Amsterdam, 1981.

\bibitem[ChW]{Chung Williams}
K.L. Chung, R.J. Williams,
{\sl Introduction to Stochastic Integration},
Birkh\"auser, Boston 1983.

\bibitem[Bar]{Bartlett}
M.S. Bartlett, {\sl An Introduction to stochastic processes},
Cambridge University Press 1962.

\bibitem[Ram]{Ramakrishnan}
A. Ramakrishnan, Proc. Cam. Phil. Soc. {\bf 46}, 595--602 (1950),
{\bf 48}, 451--456 (1952), {\bf 49}, 473--485 (1953).

\bibitem[WBKM]{WeBuKuMa} T.~Wellens, A.~Buchleitner and B.~K\"ummerer,
             H.~Maassen,
             Phys. Rev. Letters 85 (2000), 3361.



\end{references}
\end{document}